\documentstyle[aps,prl,multicol,epsfig]{revtex}
\begin{document}
\draft
\title{
First-Principles Calculation of the Superconducting Transition in MgB$_2$
within the Anisotropic Eliashberg Formalism
}

\author{Hyoung Joon Choi,$^1$ David Roundy,$^{1,2}$ Hong Sun,$^{1}$ Marvin L. Cohen,$^{1,2}$ and
Steven~G.~Louie$^{1,2}$}
\address{$^1$Department of Physics, University of California at Berkeley,
Berkeley, CA 94720, USA.\\
$^2$Materials Sciences Division, 
Lawrence Berkeley National Laboratory, Berkeley, CA 94720, USA.}
%\date{\today}

\maketitle

\begin{abstract}
We present a study of the superconducting transition 
in MgB$_2$ using the {\em ab-initio} pseudopotential density 
functional method and the fully anisotropic Eliashberg equation. 
Our study shows that the anisotropic Eliashberg equation, constructed
with {\em ab-initio} calculated momentum-dependent 
electron-phonon interaction and 
anharmonic phonon frequencies, yields 
an average electron-phonon coupling constant $\lambda = 0.61$, 
a transition temperature $T_c = 39 K$, and a boron isotope-effect exponent
$\alpha_B = 0.31$ with a reasonable assumption of $\mu^* = 0.12$.
The calculated values for $T_c$, $\lambda$, and $\alpha_B$
are in excellent agreement with transport, specific heat, and 
isotope effect measurements respectively.
The individual values of the electron-phonon coupling 
$\lambda(\vec{k},\vec{k'})$ on the various pieces of the Fermi surface
however vary from 0.1 to 2.5.
The observed $T_c$ is a result of both
the raising effect of anisotropy in the electron-phonon couplings
and the lowering effect of anharmonicity in the relevant phonon modes.
\end{abstract}
%\pacs{74.25.Kc, 63.20.Ry, 74.20.-z}

\begin{multicols}{2}[]\narrowtext

Although MgB$_2$ is a readily available $sp$-bonded
material, superconductivity in this material with a transition temperature of
$T_c$ = 39 K was found only very recently\cite{nagamatsu01}.
This relatively high $T_c$ has motivated many studies, as has
the observation that 
the detailed superconducting properties of MgB$_2$ show 
significant deviations from those calculated using the standard BCS model.
The isotope effect exponent for boron $\alpha_B$
is reduced substantially from the conventional value for
$sp$ metals\cite{budko01,hinks01}, and 
the average electron-phonon coupling strength $\lambda$
obtained from specific heat measurement\cite{wang01,bouquet01} 
seems too small to justify the high $T_c$.
In addition, specific heat measurements\cite{wang01,bouquet01},
tunneling \cite{giubileo01} and 
photoemission \cite{tsuda01} spectra, and  
point-contact spectroscopy\cite{szabo01,laube01}
show low energy excitations suggesting a secondary gap.
Theoretical calculations show that the Fermi surface 
has several pieces and is very anisotropic\cite{kortus01}, and that
the electron-phonon coupling is dominated by the in-plane B--B stretching modes
($E_{2g}$)\cite{kortus01,an01,bohnen01} which have a 
large anharmonicity\cite{yildirim01,liu01}. 
The electron-phonon interaction  varies strongly on the
Fermi surface\cite{liu01,kong01}, and a two-band model suggests 
a multigap scenario\cite{liu01}.
However, there has not yet been a quantitative, first-principles 
calculation of $T_c$ 
including the full variation 
of the electron-phonon interaction on the Fermi surface
and the anharmonicity of the phonons
to help confirm the phonon-mediating pairing mechanism 
for superconductivity in MgB$_2$.

In this letter, we present $T_c$
and isotope-effect exponents for MgB$_2$ 
obtained by solving 
the $\vec{k}$ and $\omega$ dependent Eliashberg equation.
It is shown that 
the anisotropy (i.e., the electronic-state dependence) 
of the electron-phonon interaction on the Fermi surface 
is strong enough to raise $T_c$ to 39K
even though the interaction is weakened by the anharmonicity of the phonons
as compared to the harmonic case.
In addition, it is shown that the anharmonicity of the phonons 
reduces $\alpha_B$ to 0.31.
These results show that 
conventional phonon-mediated electron pairing theory can
explain superconductivity in MgB$_2$ when both
the anisotropy of the electron-phonon interaction and the anharmonicity 
of the phonons are properly taken into account.
The solution of the full Eliashberg equation 
at low $T$  further yields different
gap values for the different parts of the Fermi surface. The gap value
distribution clusters into two groups -- a small value of $\sim$ 2 meV
and a large value of $\sim$ 7 meV. This feature and its physical
consequences will be described in more details 
in a future publication\cite{choi01}.

The phonon frequencies and electron-phonon matrix elements are
calculated  using {\em ab-initio} pseudopotentials and
the local density approximation.
We used a $18\times18\times12$ k-point grid in the Brillouin zone (BZ),
and included planewaves up to 60 Ry as a basis 
to expand the electronic wavefunctions.
The calculated equilibrium lattice constants are $a = 3.071${\AA}
and $c = 3.578${\AA}, in good agreement with measured 
values\cite{nagamatsu01}.
We performed
total energy calculations with frozen phonons for all phonon modes 
at all the high symmetry points of the BZ.
The variation  of
the total energy with a frozen phonon amplitude is fitted 
with a fourth order polynomial
to account for the phonon anharmonicity.
To obtain the harmonic phonon frequency, we use the quadratic term of the fitted
curve and calculate the frequency classically; whereas 
for the anharmonic phonon frequency, we calculate quantum-mechanical 
vibrational states including the anharmonic terms, and take the energy 
difference of the two lowest states.
In the case of the degenerate, in-plane B--B stretching modes ($E_{2g}$)
at $\Gamma$ and $A$, we calculate quantum-mechanical vibrational states in two dimensions
after the total energy is fitted in a plane
with $E(r,\theta) = E_0 + c_2r^2+c_4 r^4+(c_3 r^3 + c_5 r^5) 
\cos(3\theta)$.
We use natural atomic weights for B and Mg, that is, 
10.81 for B and 24.31 for Mg, but
$^{10}$B or $^{26}$Mg are used when we recalculate the phonon frequency 
for the isotope effect.
The linear electron-phonon matrix elements are evaluated directly 
from the total self-consistent change in the 
crystal potential caused by a frozen phonon.

Table I shows the frequency of the in-plane
B--B stretching mode ($E_{2g}$) at $\Gamma$.
This mode is doubly degenerate along the line from $\Gamma$ to $A$,
and has a large anharmonicity and
a large electron-phonon coupling.
Anharmonicity increases the frequency 
by 20\% and weakens the corresponding electron-phonon couplings 
by 30\%.
The calculated anharmonic frequency, 75.9 meV, for the $E_{2g}$ mode at $\Gamma$
agrees very well with the results from Raman measurements (75.9 meV\cite{hlinka01} and 76.9 meV\cite{goncharov01}) as well as  other theoretical
calculations\cite{yildirim01,liu01}.
The $E_{2g}$ modes at $M$, $L$, $K$, and $H$
however have very little anharmonicity and
small electron-phonon coupling. 
The strong anharmonicity and the large electron-phonon coupling 
are thus confined to phonons in 
a small volume in k-space near the $\Gamma$ to $A$ line.

The calculated phonon frequencies and electron-phonon matrix elements 
$g_{\vec{k},\vec{k}'}^j = \langle\vec{k}|\delta V_{\vec{q}}^j|\vec{k}'\rangle$
for the $j$th phonon mode
are interpolated 
onto a $18\times18\times12$ grid in the BZ
through the following three-step process.
First, we interpolate the dynamical matrices using a weighted
average of those at the symmetry points and 
obtain the phonon frequencies and eigenvectors on the grid
by diagonalizing the dynamical matrices.
Second, we interpolate the induced crystal potential change by a phonon 
on the grid  from 
the calculated crystal potential changes  
at the symmetry points using weighting factors determined from the
phonon frequencies and polarization vectors calculated on the fine grid.
Finally, we {\em calculate} the electron-phonon matrix elements on the grid 
using the interpolated crystal potential change.
All calculations are done twice for comparison: one with harmonic phonon
frequencies and another with anharmonic phonon frequencies.
To study the isotope effect, we repeat the entire procedure
with an isotopic atomic mass.

\begin{figure}
\centering
%\mbox{\epsfig{file=figure1.epsi,width=12cm,angle=0,clip=}}
\mbox{\epsfig{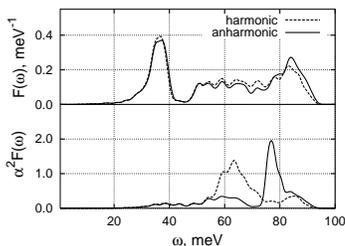}}
\caption{Phonon density of states $F(\omega)$ and the isotropic 
Eliashberg function $\alpha^2F(\omega)$ for MgB$_2$.}
\end{figure}

Figure 1 shows the phonon density of states $F(\omega)$ 
and the standard Eliashberg function $\alpha^2F(\omega)$.
The phonon density of states shows a large peak at 37 meV arising from
the van Hove singularities in the acoustic phonons,
but these phonons make no significant contribution
to $\alpha^2F(\omega)$.  There is a large dominant peak in
$\alpha^2F(\omega)$ at 63 meV for the case of harmonic phonons 
but at 77 meV for anharmonic phonons.
The dominant peak in $\alpha^2F(\omega)$ 
is caused by the in-plane B--B stretching
modes ($E_{2g}$).
Because the $E_{2g}$ modes are highly anharmonic and have very large 
electron-phonon coupling only for phonons
within a small volume along the $\Gamma$ to $A$ line in k-space,
anharmonicity has little effect on $F(\omega)$, but
it causes a big shift in $\alpha^2F(\omega)$.
In the case of harmonic phonons, as is shown in Table I, the {\em isotropic}
average electron-phonon coupling constant,
$\lambda = 2 \int d\omega \alpha^2F(\omega)/\omega$,
is 0.73 and the logarithmic average frequency, 
$\omega_{ln} = \exp[(2/\lambda) \int 
d\omega\alpha^2F(\omega)\ln\omega/\omega]$, is 59.4 meV.
These values and the overall shape of $\alpha^2F(\omega)$ without anharmonicity
in the present calculation
are in good agreement with previous calculations\cite{bohnen01,liu01}. 
With anharmonicity, $\lambda$ is reduced to 0.61 and $\omega_{ln}$ 
is increased to 63.5 meV. 
Since $\lambda$ corresponds to the 
mass enhancement factor for the density of states 
at the Fermi level regardless of anisotropy in the electron-phonon
interaction\cite{allen82}, we can compare the calculated $\lambda$ with
results of specific heat measurements.
The reduced value of  $\lambda = 0.61$ due to anharmonicity 
agrees very well with result of specific heat measurements 
which give a $\lambda$
of 0.58\cite{wang01} and 0.62\cite{bouquet01}.
This agreement is evidence that phonon anharmonicity weakens
the electron-phonon interaction in MgB$_2$.
However if this value of $\lambda=0.61$ is used in the McMillan\cite{mcmillan68}
or the Allen-Dynes\cite{allen75} formula for $T_c$, the predicted
$T_c$ would be far lower than experiment.

Unlike previous studies,
we solve the fully anisotropic Eliashberg equation for superconductivity
in MgB$_2$.
The anisotropic Eliashberg equation at $T_c$\cite{allen82} is
\begin{eqnarray}
\nonumber Z(\vec{k},i\omega_n) & = & 1+f_n s_n\sum_{\vec{k}'n'}W_{\vec{k}'}\lambda(\vec{k},\vec{k}',n-n') s_{n'} \\
\nonumber Z(\vec{k},i\omega_n)\Delta(\vec{k},i\omega_n) & = & 
\sum_{\vec{k}'n'}W_{\vec{k}'}f_{n'}\left[\lambda(\vec{k},\vec{k}',n-n')-\mu^*\right] \\
 & &
~~~~\times \Delta(\vec{k}',i\omega_{n'})~,
\end{eqnarray}
where $\omega_n = (2n+1)\pi T_c$, $f_n = 1/|2n+1|$, and
$W_{\vec{k}}$ is the fraction
of the density of states at $\vec{k}$ on the Fermi surface.
For the definition of $Z$, $\Delta$, $\lambda(\vec{k},\vec{k}',n)$, and $s_n$,
see Ref. \cite{allen82}.
With the exception of $\mu^*$, our calculation
of the phonon frequencies and electron-phonon interaction provides
all the material parameters for solving Eq. (1) 
and hence for obtaining $T_c$ from first principles.
The dimensionless Coulomb pseudopotential, $\mu^*$\cite{morel62},
is the only free parameter in our calculation; but it is known to be
of order 0.1 in most metals\cite{mcmillan68,morel62,carbotte90} 
and we show below that the superconducting properties of MgB$_2$ are 
insensitive to $\mu^*$.
For comparison, we also calculate $T_c$ using
the isotropic Eliashberg equation, 
\begin{eqnarray}
\nonumber Z(i\omega_n) & = & 1+f_n s_n\sum_{n'}\lambda(n-n')s_{n'} \\
Z(i\omega_n)\Delta(i\omega_n) & = & 
\sum_{n'}f_{n'}\left[\lambda(n-n')-\mu^*\right]\Delta(i\omega_{n'})~,
\end{eqnarray}
where 
$\lambda(n) \equiv \sum_{\vec{k} \vec{k}'} W_{\vec{k}} W_{\vec{k}'}
\lambda(\vec{k},\vec{k}',n)$.
Hence $\lambda(n)$ is the electron-phonon coupling averaged over all
pairs of $(\vec{k},\vec{k}')$ on the Fermi surface.
($\lambda(n\!=\!0)$ is equal to the specific heat $\lambda$ discussed above.)
The isotropic Eliashberg equation is thus a special limited case of the
more general anisotropic equation. 
If the electron-phonon interaction $\lambda(\vec{k},\vec{k}',n)$ did not
depend strongly on the electronic states on the Fermi surface,
the isotropic equation would be an appropriate approximation.

\begin{figure}
\centering
%\mbox{\epsfig{file=figure2.epsi,width=12cm,angle=0,clip=}}
\mbox{\epsfig{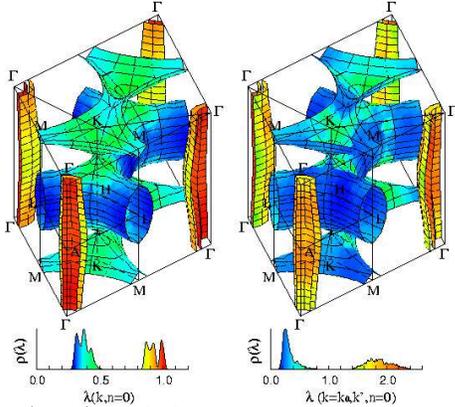}}
\caption{(color) Variation of the electron-phonon interaction $\lambda$ on
the Fermi surface of MgB$_2$.
The left plot shows the mass enhancement factor given by
$\lambda(\vec{k},n\!=\!0)$.
The right plot shows $\lambda(\vec{k}=\vec{k}_0,\vec{k}',n\!=\!0)$ 
as a function of $\vec{k}'$
for a fixed $\vec{k}_0$ on the Fermi surface near $\Gamma$.}
\end{figure}

\begin{figure}
\centering
%\mbox{\epsfig{file=figure3.epsi,width=12cm,angle=0,clip=}}
\mbox{\epsfig{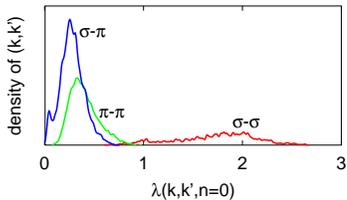}}
\caption{(color) Number density of $(k,k')$ pairs on the Fermi surface
versus the value of $\lambda(\vec{k},\vec{k}',n\!=\!0)$.
The number density is split into three sets:
both $k$ and $k'$ are 
on the two $\sigma$ cylindrical sheets of the Fermi surface (red line),
both on the two $\pi$ tubular sheets (green line),
and one on a $\sigma$ cylindrical sheet 
and the other on a $\pi$ tubular sheet (blue line).}
\end{figure}

Figure 2 shows the variation of the calculated electron-phonon interaction 
on the Fermi surface of MgB$_2$. 
The Fermi surface of MgB$_2$ consists of four
sheets; two holelike coaxial cylinders consisting of 
$\sigma$-boron in-plane states along $\Gamma$ to $A$,
a holelike tubular network  of $\pi$-boron states
connecting regions near $K$ and $M$, and 
an electronlike tubular network of $\pi$-boron states
connecting regions near $H$ and $L$.
In Fig. 2, the mass enhancement factor for states at $\vec{k}$ given by
$\lambda(\vec{k},n\!=\!0) =  
\sum_{\vec{k}'}W_{\vec{k}'}\lambda(\vec{k},\vec{k}',n\!=\!0)$ 
shows two well-separated sets of values on the Fermi surface.
$\lambda(\vec{k},n\!=\!0)$ is about 0.8 $\sim$ 1.0
on the two $\sigma$ cylindrical sheets and is only about 0.3 $\sim$ 0.5
on the two $\pi$ sheets.
For more detail,
we depict the value of $\lambda(\vec{k}=\vec{k}_0,\vec{k}',n\!=\!0)$ 
as a function of $\vec{k}'$ 
for a fixed $\vec{k} = \vec{k}_0$ on the Fermi surface near $\Gamma$.
It shows strong and varying strength for scattering onto the $\sigma$
cylindrical surfaces but rather weak strength for scattering 
to the $\pi$ sheets.
Figure 3 shows the number density of $(\vec{k},\vec{k}')$ pairs on the Fermi
surface plotted as a function of the value of $\lambda(\vec{k},\vec{k}',n\!=\!0)$.
The coupling strength $\lambda(\vec{k},\vec{k}',n\!=\!0)$ 
between states on the $\sigma$ cylindrical sheets has values exceeding 2.0
which is much larger than those within the $\pi$ tubular sheets
or between a $\sigma$ cylindrical sheet and a $\pi$ tubular sheet.
All this information and $\lambda(\vec{k},\vec{k}',n)$ with non-zero $n$
are taken into account in solving the anisotropic Eliashberg equation.

The anisotropic Eliashberg equation including anharmonicity in the phonon
frequencies
yields 42K $\geq T_c \geq$ 37K for $0.10 \leq \mu^* \leq 0.14$.
In particular,  $T_c$ is 39K when $\mu^* = 0.12$.
To investigate the role of anisotropy in the
electron-phonon interaction and of anharmonicity of the phonons,
we calculate $T_c$ disregarding one or the other, as shown in Table I.
If we neglect the anisotropy and calculate $T_c$ with the isotropic Eliashberg
equation, $T_c$ drops to 19K for $\mu^* = 0.12$. 
This shows that the strong variation in the 
electron-phonon coupling of scattering on the Fermi surface is 
crucial to the observed high $T_c$ in MgB$_2$.
As another comparison, if we calculate $T_c$ using the anisotropic Eliashberg 
equation 
but neglecting the anharmonic effect in the phonon frequencies,
$T_c$ goes up to 55K  for $\mu^* = 0.12$.
Hence anharmonicity lowers $T_c$ in MgB$_2$.
Thus we conclude that anisotropy in MgB$_2$ is essential to
produce the anomalously high $T_c$, especially in view of the fact that
the electron-phonon interaction is weakened by anharmonicity.
We note that in MgB$_2$, an average electron-phonon 
coupling $\lambda$ cannot be correctly
determined from $T_c$ using 
the McMillan\cite{mcmillan68,mcm} or Allen-Dynes equations\cite{allen75},
however a determination of $\lambda$ from the specific heat measurement
is still valid.
The $T_c$ of MgB$_2$ is not a function 
of the usual isotropically averaged electron-phonon interaction 
$\lambda$ given above;
it depends on the 
details of electron-phonon interactions on the full Fermi surface.
This explains the apparent discrepancy between the values of
$\lambda$ estimated
from  specific heat measurements and $\lambda$ estimated from $T_c$
using simplied isotropic models.

To calculate the isotope-effect exponent $\alpha$
($T_c \propto M^{-\alpha}$), we recalculate $T_c$
using the mass of either
$^{10}$B or $^{26}$Mg in place of the natural atomic weight.
Since the relevant phonons are related to motion of the B atoms,
the Coulomb pseudopotential $\mu^*$ also depends on the atomic mass of boron $M_B$.
The change of $\mu^*$ is simply
\begin{equation}
\delta\mu^* = -(\mu^*)^2 \frac{\delta M_B}{2M_B}~,
\end{equation}
using
$ \mu^* = \mu/[1+\mu \ln (\epsilon_F/\omega_0)] $
and $\omega_0 \propto M_B^{-0.5}$.

Table I shows calculated isotope-effect exponents with $\mu^* = 0.12$  
both with and without phonon anharmonicity.
Without anharmonicity, the slight deviation of
the sum of the two exponents, $\alpha_B$ and $\alpha_{Mg}$, from
the value of 1/2 is due to the change of $\mu^*$ given by Eq. (3). If we neglect
$\delta\mu^*$, the total isotope-effect exponent would be 0.5.
In contrast, when we include anharmonicity in the
phonon frequency, the isotope-effect exponent for boron is
substantially suppressed.
We obtain $\alpha_B = 0.31$ and $\alpha_{Mg} = 0.05$ from the anisotropic 
Eliashberg equation with anharmonic phonon frequencies. 
In this case,
the contribution of $\delta\mu^*$ to the decrease of $\alpha_B$ is only 0.02,
so the anomalously low isotope-effect exponent 
is primarily due to phonon anharmonicity.

In conclusion, we have shown from first-principles calculations that MgB$_2$
is a conventional phonon-mediated superconductor whose properties require,
for a correct description, a solution of the fully
anisotropic Eliashberg equation 
including phonon anharmonicity.
The isotropic Eliashberg equation seriously
underestimates $T_c$ because it fails to
account for the $(\vec{k},\vec{k}')$-dependency  
of the electron-phonon interaction on the Fermi surface\cite{mcm}.
We show that the electron-phonon coupling is exceedingly strong for
certain pairs of $(\vec{k},\vec{k}')$ on the disconnected Fermi surface
of this material.
The anisotropy of the electron-phonon interaction in MgB$_2$ 
is strong enough to
produce the observed $T_c$ of 39 K in spite of a moderate 
average electron-phonon interaction as also seen in specific heat
measurements.  In addition,
we have shown that the anharmonicity of the phonons in MgB$_2$ weakens 
the electron-phonon interaction and 
reduces the boron isotope-effect exponent.

This work was supported by National Science Foundation Grant No. DMR00-87088,
and by the Director, Office of Science, Office of Basic Energy Sciences
of the U. S. Department of Energy under Contract DE-AC03-76F0098.
Computational resources
have been provided by the National Science Foundation at the National Center
for Supercomputing Applications and by the National Energy Research Scientific
Computing Center. 
Authors also acknowledge financial support from the Miller Institute
(H.J.C.) and from the Berkeley Scholar Program funded by the Tang Family 
Foundation (H.S.).

\begin{table}
\caption{Transition temperature $T_c$ and  isotope-effect exponents $\alpha$
with $\mu^* = 0.12$.
Numbers in parentheses %in the row for $\alpha_B$ 
are the values of  $\alpha_B$ when  $\delta\mu^*$  of Eq.(3) is ignored. 
The averaged electron-phonon coupling $\lambda$ and 
the frequency $\omega_{ph}$ of 
the in-plane B--B stretching modes ($E_{2g}$) at $\Gamma$ are also included.
}
\begin{tabular}{cccccc}
               & \multicolumn{2}{c}{harmonic}            & \multicolumn{2}{c}{anharmonic}              &  experiment\\
               &   isotropic & anisotropic & isotropic    & anisotropic  &   \\
\hline
 $T_c$         &    28 K      & 55 K     &     19 K      &    39 K     & 39 K\cite{nagamatsu01}  \\
 $\alpha_B$    &  0.41  &  0.48  & 0.21  &  0.31 & 0.26\cite{budko01},0.30\cite{hinks01} \\
               &  (0.45) &  (0.49) & (0.27) &  (0.33) & \\
 $\alpha_{Mg}$ &    0.04      &   0.02      &  0.06      &  0.05      & 0.02\cite{hinks01} \\
\hline
 $\lambda$      &   \multicolumn{2}{c}{0.73}     &    \multicolumn{2}{c}{0.61}      & 0.58\cite{wang01}, 0.62\cite{bouquet01} \\
 $\omega_{ph} $  &   \multicolumn{2}{c}{62.7 meV} & \multicolumn{2}{c}{75.9 meV}  &75.9\cite{hlinka01}, 76.9\cite{goncharov01} 
\end{tabular}
\end{table}

\end{multicols}


\begin{thebibliography}{99}
\bibitem{mcm} We have also tested the widely used McMillan
formula for $T_c$\cite{mcmillan68}. 
In MgB$_2$, the McMillan formula underestimates $T_c$ by typically 30\%
even when compared with results from the isotropic Eliashberg equation.
\end{thebibliography}
\end{document}